\documentstyle[psfig,12pt,aaspp4,flushrt]{article}  
\def\be{\begin{equation}}
\def\ee{\end{equation}}
\def\etal{{\it et al. }}
\def\kms{km~s$^{-1}$~}

\def\smalltype{\let\rm=\eightrm \let\bf=\eightbf
  \let\it=\eightit \let\sl=\eightsl \let\mus=\eightmus
  \baselineskip=9.5pt minus .75pt
  \rm}

\begin{document}

\title{Galaxy Peculiar Velocities and Infall onto Groups}

\author{M. L. Ceccarelli, C. Valotto, D. G. Lambas}
\affil{IATE, Observatorio Astron\'omico,
Universidad Nacional de C\'ordoba, Laprida 854, C\'ordoba 5000, Argentina; laura@mail.oac.uncor.edu}
\author{N. Padilla}
\affil{Departamento de Astronom\'{\i}a y Astrof\'{\i}sica, Pontificia Universidad Cat\'olica
de Chile, Vicu\~na Mackenna 4860, Santiago, 22, Chile.}
\author{R. Giovanelli and M. Haynes}
\affil{Department of Astronomy, Space Sciences Building, Cornell University,
Ithaca, NY 14853 }

\begin{abstract}

We  perform  statistical analyses to study the 
infall of galaxies onto groups and
clusters in the nearby Universe.
The study is based on the Updated Zwicky Catalog and Southern Sky Redshift Survey 2 group catalogs and
peculiar velocity samples.
We find a clear signature of infall of galaxies onto groups over a wide
range of scales $5$ h$^{-1}$ Mpc$<$ r $< 30 $ h$^{-1}$ 
Mpc, with an infall amplitude on 
the order of a few hundred kilometers per second.
We obtain a significant increase in the infall amplitude with 
group  virial mass ($M_{V}$) and  
luminosity of group member galaxies ($L_{g}$).
 Groups with $M_{V} < 10^{13}$ M$_{\odot}$ show infall velocities
$V_{infall} \simeq 150$ km s$^{-1}$
 whereas for  $M_{V} > 10^{13}$ M$_{\odot}$ a larger infall is observed,
$V_{infall} \simeq  200$ km s$^{-1}$.
Similarly, we find that galaxies surrounding 
groups with  $ L_{g}< 10^{15}$ L$_{\odot}$ have $V_{infall} \simeq  100$ km s$^{-1}$, 
whereas for $L_{g} > 10^{15}$ L$_{\odot}$ groups,
the amplitude of the galaxy infall can be 
as large as $V_{infall} \simeq 250$ km s$^{-1}$.
The observational results are compared with the
results obtained from mock group and galaxy samples 
constructed from numerical simulations,
which include galaxy formation through semianalytical models.
We obtain a general agreement between the results from the mock catalogs and 
the observations. The infall of galaxies onto groups is suitably reproduced 
in the simulations and, as in the observations, larger virial 
mass and luminosity groups exhibit the largest galaxy infall amplitudes. 
We derive estimates of the integrated mass overdensities associated with 
groups by applying linear theory to the infall velocities after correcting for 
the effects of distance uncertainties obtained using the mock catalogs.
The resulting overdensities are consistent with a power law with $\delta \sim 1$
at $r\sim 10 h^{-1}Mpc$. 

\end{abstract}


\section{Introduction}
Inhomogeneities in the distribution of matter
are the source of  
peculiar velocities in the universe. The nature of this 
velocity field depends on the
local density. High density regions show a collection of random motions
typical of virialized objects whereas,  
low density environments, on the other hand, are more likely to show
streaming motions: objects falling towards larger potential wells
constantly increasing the amplitude of their clustering strength (Diaferio \& Geller, 1997).

A possible model for describing the dynamical  behavior of objects near 
regions of high density contrast is provided by the spherical 
infall model (for a discussion see Tolman, 1934; Gunn \& Gott, 1972; Silk, 1974; Elkolm \& Teerrikorpi, 1994).
This model simplifies the problem by assuming that the initial density perturbation 
responsible for the formation of an object
is spherically symmetric.  The gravitational field around this perturbation
traces the symmetry of the matter, and its pull induces peculiar motions
over the surrounding area.  Here the velocity field takes the form of
a collapsing streaming motion towards the local density maximum.
In the linear perturbation case, the infall velocity ($V_{infall}$)
depends on the distance to the local density maximum and is directly 
related to the density contrast (Peebles, 1980),

\begin{equation}
{V}_{infall}=-(1/3) \Omega_{0}^{0.6} H_{0} r \delta (r)
\end{equation}

In high density regions, as is the case of
neighborhoods of clusters and galaxy groups, 
linear theory is not expected to provide an adequate description
(Croft et al. 1999). However, 
if we assume that mass overdensities are spherically symmetric, 
an approximate solution for the non-linear collapse can be found. 
This solution treats the overdensity as 
an isolated Friedmman universe with its own value of 
$\Omega_{0}$ (for more details see Gunn \& Gott, 1972; Bondi, 1974; 
Silk, 1974; Gunn, 1978; R\"egos \& Geller, 1989 and references therein).
For these non-linear regions, the relation for linear collapse 
is replaced by an approximate solution (Yahil, 1985).
This solution has shown to give
accurate results outside the virialized regions of
groups and clusters of galaxies.

In this work we investigate the structure of the peculiar velocity field around 
groups of galaxies in the nearby universe.   
We apply the spherical infall model to test the effects of overdensities in the
density field, traced by groups of galaxies, on the dynamical properties of 
their surrounding regions.

This paper is organized as follows.
In section 2 we present the observational
data samples used to perform the statistical analyses. 
We describe the semi-analytic mock catalogs in section 3.
Section 4 contains the description of the statistical methods adopted
to obtain the results shown in sections 5 and 6. 
In section 7 we show the results of
the method described in section 3 applied to the mock catalogs.
In section 8 we apply a correction factor to consider distance uncertainties on
the amplitude of infall velocities and we use these estimates to predict
the integrated mass overdensity associated to groups from linear theory. 
Finally, we make a brief discussion of the results.

\section{Observational Data}

In this section we describe in detail the observational samples used in this
paper. We use two different sets of data.  The first set is composed of
tracers of the density peaks, which in this work are galaxy groups 
from the Updated Zwicky Catalog (UZC; Merch\'an et al. 2000) and the Southern 
Sky Redshift Survey (SSRS2).  
The second set of data contains galaxy peculiar velocity 
information taken from the compilation of Giovanelli and Haynes (2002, 
hereafter CPV), which we use to investigate the velocity field around 
mass concentrations.

We use the UZC group catalog constructed by Merch\'an et.al (2000).
These groups were identified from the UZC galaxy catalog
(Falco et al. 1999)
which contains 19,369 galaxies with apparent Zwicky magnitudes
$m_{Zw} \le 15.5$ and
with a $96\%$ completeness in redshift. The region covered by the catalog is bounded by
$20^h\leq \alpha_{1950}\leq 4^h$,  $8^h\leq \alpha_{1950}\leq 17^h$
and $-2.5^o\leq \delta_{1950}\leq 90^o$, providing accurate coordinates
within 2$"$ and reliable redshifts in the range $cz=0-25,000$ km s$^{-1}$, with a
reasonably complete sky coverage
(see Falco et al. 1999 for more details). 
The algorithm adopted by Merch\'an et al. (2000)
for the construction of the group catalog
follows the basic procedure described by Huchra \& Geller (1982) with the
improvements introduced by Maia et al. (1989) and Ramella et al. (1997) that minimize the
number of interlopers. The group inner regions have a density contrast
relative to  the mean density of galaxies 
$\delta\rho/\rho=80$.
The UZC group catalog (GUZC) contains systems with at least four members and
mean radial velocities, $V_{gr} \le$ 15,000 \kms 
comprising a total number of 513 groups.

\begin{figure}
\plotone{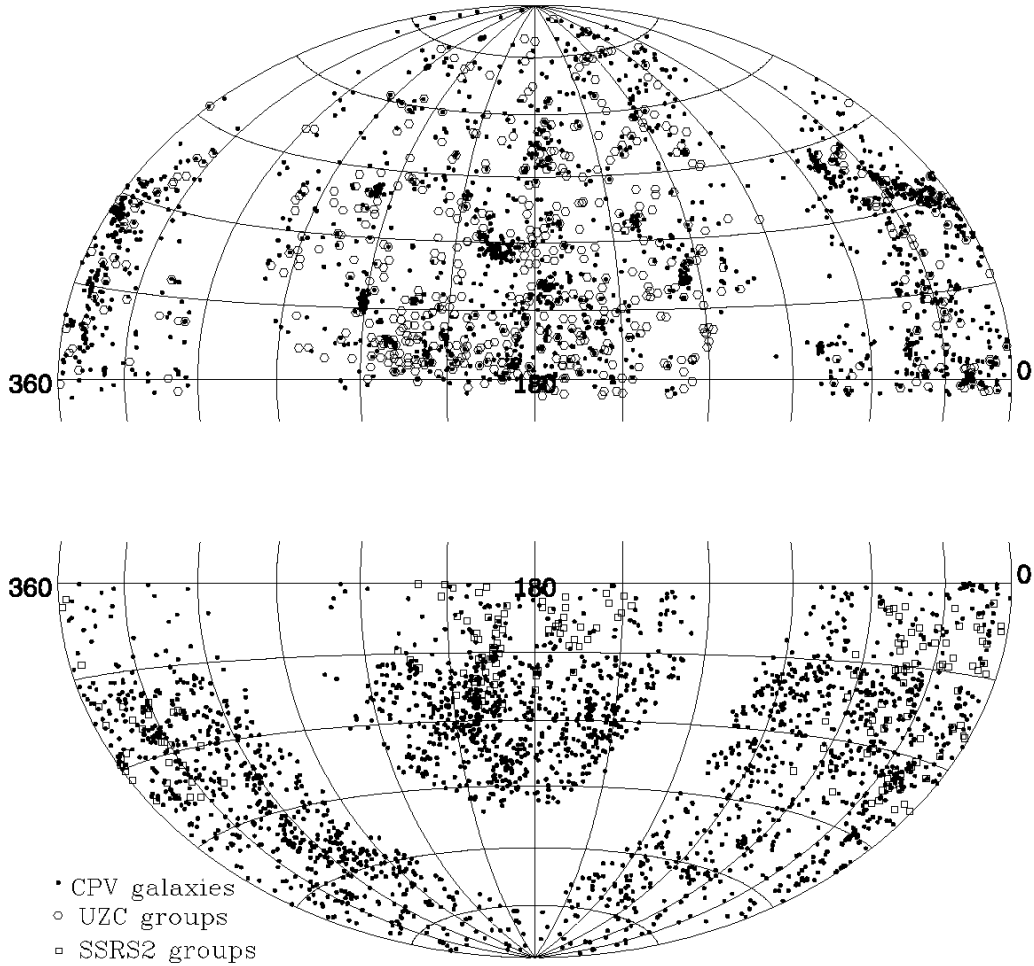}
\caption{Top: Angular distribution of 
north CPV galaxies (points) and UZC groups (hexagons) for the northern 
hemisphere in equatorial coordinates. 
Bottom: Angular distribution of south CPV galaxies (points) and 
SSRS2 groups (open squares) for the southern hemisphere, also in 
equatorial coordinates} 
\label{fig:uzcg}
\end{figure}

The top half of figure \ref{fig:uzcg} shows the angular distribution of the 
CPV galaxies and UZC groups.  As can be seen, these objects are all concentrated
on the northern hemisphere, with a sampling rate that does not
present important variations across the solid angle covered.

We also use groups identified from the SSRS2 catalog (Da Costa et al. 1998)
which comprises 5369 objects with apparent B magnitudes
$m_{B} \le 15.5$.  The region covered is 1.70 sr in the southern celestial 
hemisphere providing accurate coordinates within 1$"$.
This catalog of groups was constructed using the same algorithm
used for the GUZC (Merch\'an, 2000).
The SSRS2 group catalog contains systems with a minimum of four members and
mean radial velocities, $V_{gr} \le$ 15,000 \kms, 
comprising a total of 386 groups. The bottom half in figure \ref{fig:uzcg} shows 
the angular positions of groups from the SSRS2 and CPV southern galaxies.

\subsection {Peculiar velocity data}

Peculiar velocities have been derived for 4452 galaxies
using I-band photometry and either optical long slit spectroscopy
or global HI profiles (e.g. Dale \& Giovanelli 2000). The sample
includes objects in the Haynes \etal ~(1999), Mathewson \& Ford
(1996) and Giovanelli \etal ~(1997a) compilations, as well as
more recently processed data, the public presentation of which
is in preparation by our group.
Widths derived from optical rotation curves follow the method
outlined in Giovanelli \& Haynes (2002). Peculiar
velocities were derived using the template relation and the
derivation technique presented in Giovanelli \etal ~(1997b).
Analyses of the large--scale flow properties of the peculiar
velocity field based on these samples are summarized in Dale
\& Giovanelli (2000).

\section{Mock Catalogues with Semi-analytic Models of Galaxy Formation}
\label{s:mocks}

In order to compare model predictions to observational results
we use mock catalogs extracted from a high resolution N-body 
simulation populated with galaxies using the semianalytic model 
{\tt GALFORM} (Cole et al. 2000; Benson et al. 2002),
by the Theory Group at the University of Durham.  
The numerical simulation uses standard $\Lambda$CDM parameters, a normalization
$\sigma_{8}=0.80$, and a primordial spectral index of $n=0.97$, 
in agreement with the constraints from the first year of data from WMAP 
(Spergel et al. 2003).
The N-body simulation box is $250 h^{-1}$ Mpc on a side and contains 
$1.25\times10^{8}$ dark matter particles of mass 
$1.04\times10^{10}h^{-1}M_{\odot}$. 
Dark matter halos are identified in the $z=0.$ output using
a friends of friend method and rejecting those halos with less than $10$ particles.
As a result, the halo resolution limit is 
$1.04\times10^{11}h^{-1}M_{\odot}$. 
The procedure followed by the Durham group for populating the dark matter
halos follows the technique described by Benson et al. (2002). 
Essentially, the {\tt GALFORM} code is run for each halo, and galaxies are 
assigned to a subset of dark matter particles in the halo.
Around $90\%$ of central galaxies brighter than 
$M_{b_{\rm J}}-5\log_{10} h=-17.5$ 
are expected to be in halos resolved by the simulation. 
The effective limit of the catalog was extended to 
$M_{b_{\rm J}}-5\log_{10} h=-16$ using a 
separate {\tt GALFORM} calculation for 
a grid of halo masses below the resolution limit of the N-body 
simulation. These galaxies are assigned to particles that are not 
identified as part of a dark matter halo.

\begin{figure}
\plotone{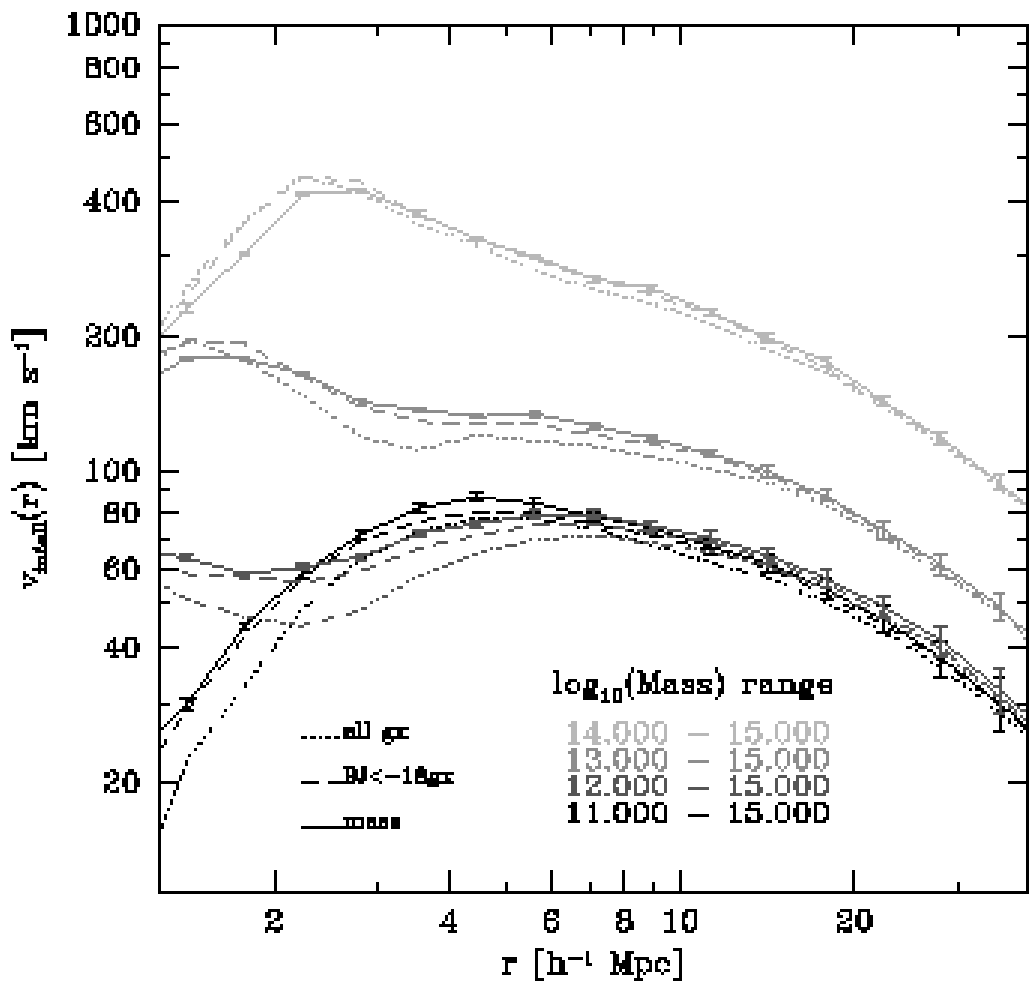}
\caption{Mean infall velocity of particles in the LCDM simulation 
(solid lines) and galaxies (all semianalytic galaxies: dotted lines;
galaxies such that $B_j<-18$: dashed lines)
as a function of radial distance to the center of mass of
the groups.}  
\label{fig:vcube20}
\end{figure}

First, we investigate the infall of mass particles onto the dark matter 
halos in the numerical simulation.
Figure \ref{fig:vcube20} shows the mean infall velocity of dark matter particles and 
semianalytic galaxies in the simulation cube 
as a function of distance to the halo center of mass.
We show results for the full sample of semianalytic galaxies
and for those with $B_j<-18$.
This figure shows in different shades of gray, 
results for subsamples of halos defined using different lower
mass limits.
As can be seen, galaxies show a similar infall velocity to dark matter particles
even when they are restricted to brighter galaxies.  The infall velocities
seem to be marginally lower for galaxies than for the dark matter.

In a second step, we extract mock catalogs from the numerical
simulation.  These play a fundamental role in our analysis.  We use
them to tailor our algorithms to extract the maximum 
streaming motion signal and to
asses random and systematic errors in our
calculations.  We estimate systematic errors by comparing the results
measured from the mock catalogs with the results from measuring
the streaming motions in the simulation cube.  Random errors
are calculated by the dispersion of streaming motions measured
from each individual mock galaxy around the mean infall velocities from
all the mock galaxies.  

\begin{figure}
\plotone{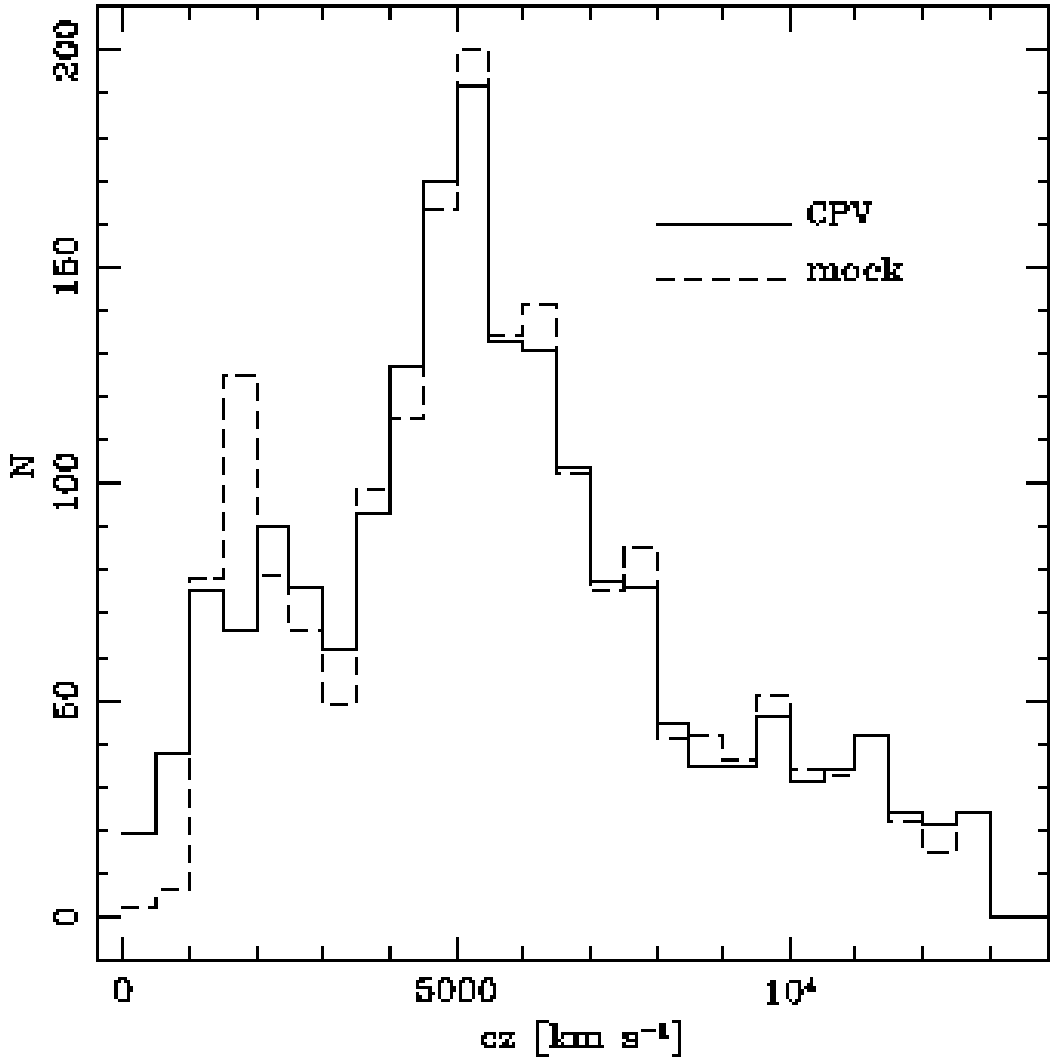}
\caption{Radial velocity  distribution of 
CPV galaxies (solid histogram) and mock CPV galaxies (dashed histogram)}
\label{fig:nzn}
\end{figure}

Mock UZC and SSRS2 are constructed by placing 
an observer at a random 
location within the simulation box and applying the radial and 
angular selection function of the UZC and SSRS2, respectively.
We also assign each mock galaxy a morphological type, which
is assumed to be a function of the stellar formation rate (SFR;
high SFR corresponds to late types, low SFR to early types).
The group finding algorithm is applied to the galaxy mock catalog
in the same way as for the real data to construct a mock group catalog
that mimics UZC groups.  In order to obtain a CPV mock catalog we 
take account the fact that CPV data comprise mainly late-type galaxies, so 
we apply a further constraint to account for this in our mock galaxy 
sample by restricting it to late types. Finally, we also impose 
the radial velocity distribution of CPV galaxies to this sample 
(see figure \ref{fig:nzn}).

In addition, the mock catalogs are selected so as to reproduce
the local group velocity with respect to the cosmic microwave background and also the
surrounding structures, which consist mainly of reasonably important
filaments aligned with the observer's line of sight to $\approx 30$ degrees.

Figure \ref{fig:nzn} shows the number of CPV
galaxies as a function of redshift [$N(z)$].
In the figure, the mock redshift distribution is shown as a dashed histogram and 
the real data of $N(z)$ are shown as a solid histogram.

\section{Statistical Methods}

In this work, we analyse the statistical properties of galaxy peculiar 
velocities in the neighborhoods of galaxy groups.
Galaxies are considered to be
test particles under the gravitational 
action of a spherically symmetric overdensity in the matter distribution,
traced by the positions of galaxy groups. 

The infall model provides
a description for the radial collapse of galaxies towards
higher density regions.    
We expect this model to be a good approximation to the
actual dynamics in the local universe, with a less important
tangential component in the streaming motions of galaxies.
In this analysis we will assume the motions are only radial.

We can find the mean infall of galaxies onto groups and clusters
using galaxy peculiar velocities, which give only the line-of-sight
projection of the three-dimensional velocity vector.
The effect of this projection can be described in a simple way as a function of
the galaxy position relative to both the group center and the observer
by the linear relation 
\begin{equation}
V_{r}(r,\theta)=V_{infall}(r) \cos(\theta),
\label{eq:v}
\end{equation}

\noindent
where $\theta$ is the angle subtended by the observer and the galaxy
as seen from the group center. We have neglected in this analysis group 
peculiar velocities whose effects are expected to cancel out in this 
statistical study.\\

We analyse  the dependence  
of galaxy peculiar velocities, $V_{pr}$, on 
$\cos(\theta)$ in different spherical concentric 
shells around the individual groups 
and calculate averages $<V_{pr}(\theta)>$ and its standard mean deviation 
$\sigma_{V_{pr}}$ 
for different bins in $\cos(\theta)$ for the total group sample.

In order to enhance the statistical significance of our
results we adopt an iterative process by which we repeatedly
calculate $<V_{pr}>$ from a reduced sample of galaxies.
The reduced sample is constructed by removing all the galaxies which
lie at more than $2\sigma_{V_{pr}}$ from the mean linear relation calculated
from the sample of galaxies used in the previous iteration.
We adopt four iterations, which is enough to stabilize the results. 

The mean infall amplitude $V_{infall}(r)$ as a function of scale $r$ is
derived from least-squares linear fitting applied to the observed $<V_{pr}>$ versus $cos \theta$
measured on shells of radius $r$ around the groups. Then $V_{infall}(r)$ is simply the slope
of the line fitted to the data (see equation \ref{eq:v}).

\section {Analysis and Results}

In order to analyze the dynamical properties of CPV galaxies we
have selected groups in the radial velocity range 
$2,000$ km s$^{-1}< cz < 8,000$ km s$^{-1}$. 
We define concentric spherical shells around each group
in our samples and identify the galaxies that lie in them.
Once we do this, we can calculate the angle $\theta$ subtended
by the directions to the galaxy and the observer from the center
of the group.

\begin{figure}
\plotone{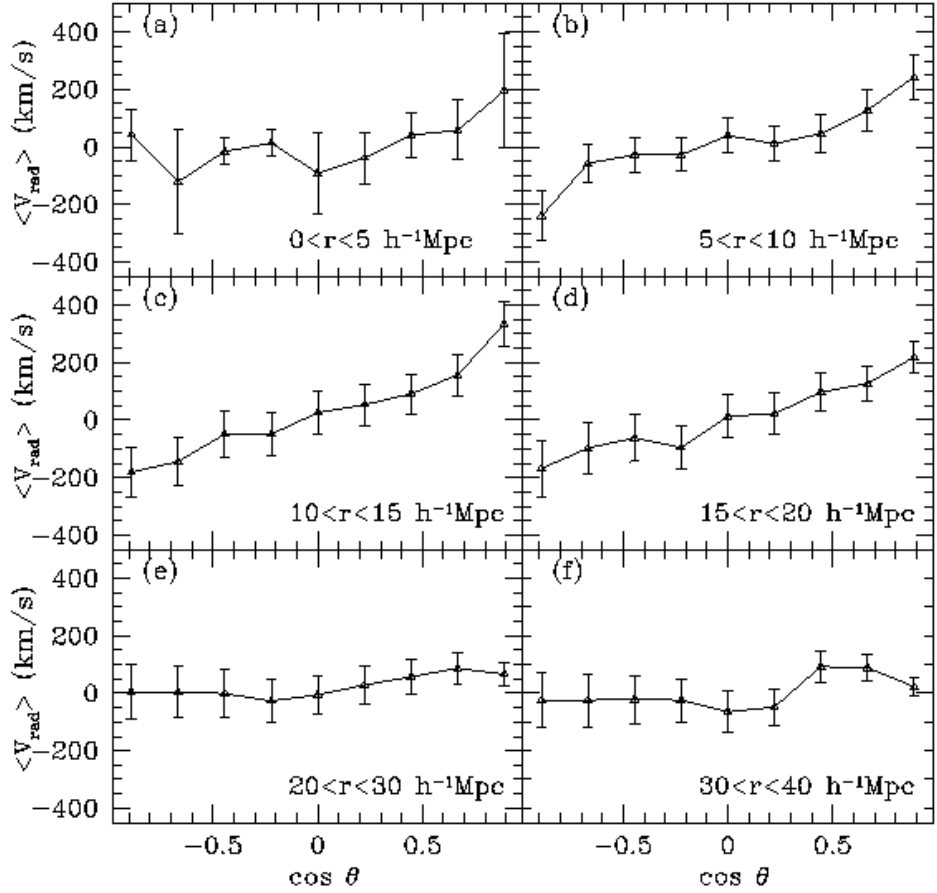}
\caption{Mean values $<V_{pr}>$  as a function of the 
angle $\theta $ subtended by the group-galaxy and
group-observer directions for the CPV galaxies and UZC groups.
The units of $r$ are h$^{-1}$ Mpc.
The different panels correspond to different distances to the group center: 
(a) $0<r<5$;
(b) $5<r<10$;
(c) $10<r<15$;
(d) $15<r<20$;
(e) $20<r<30$;
(f) $30<r<r40$}
\label{fig:vpanels}
\end{figure}

In figure \ref{fig:vpanels} we show the mean projected peculiar velocity 
as a function of $\theta$ bins for the CPV galaxies and UZC groups.
The different panels correspond to spherical shells of
different radii $r$.
As can be seen in the figure, there is an important variation 
of $V_{infall}(r)$ with shell radius.
In this figure, errors are derived from the scatter of measurements obtained
from the data.
At small $r$, the errors are larger than 
the mean peculiar velocity because galaxies do not follow radial 
trajectories near the center of the groups (panel a).
In the external regions errors become smaller and the 
infall motions show stable values over a wide range of distances.
The maximum in infall amplitude occurs at $5$ h$^{-1}$ Mpc $<$ $r$ $<$ $30\ $ h$^{-1}$ Mpc; 
at these scales the predominant motions are aligned with
the radial direction (panels b, c and d).
At scales $r>30$ h$^{-1}$ Mpc, 
the influence of surrounding structures starts to affect 
the values of infall velocities (panels e and f), which become consistent
with values $\sim 0$km s$^{-1}$, because of the large error bars.

\begin{figure}
\plotone{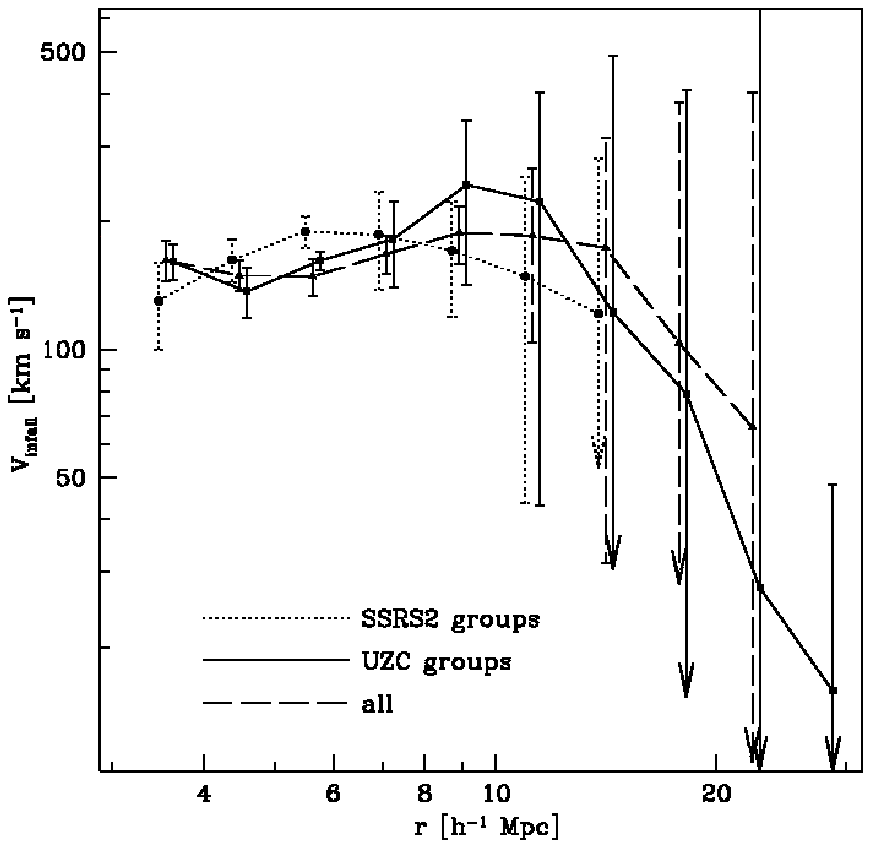}
\caption{Mean streaming motions of CPV galaxies onto UZC groups in northern 
hemisphere (solid line), SSRS2 groups in southern hemisphere (dotted line), 
and the combined data set (dashed line). Errors
are calculated using mock catalogs.}
\label{fig:vns}
\end{figure}

We repeat this analysis using groups and galaxies in the southern hemisphere
and show in figure \ref{fig:vns} the resulting values 
of streaming motions $V_{infall}$ for both
the SSRS2 and UZC galaxies (from the southern and northern hemisphere respectively)
and for the combined set of samples (north and south).
The solid line in the figure \ref{fig:vns} summarizes the results 
shown in figure \ref{fig:vpanels}. Here each point corresponds 
to the slope of the lines fitted in each panel of \ref{fig:vpanels},
and errors represent the scatter of results from individual mock catalogs
around their mean (see section 3).
As can be seen in figure \ref{fig:vns},  
there is a clear indication of infall motions outside the virialized regions.
At the largest radius the infall signal is negligible, an expected
effect caused by the action of the gravitational pull of neighboring groups.
The results are similar 
for both hemispheres.

\section{Dependence on Virial Mass and Group Luminosity}

Once we have a direct detection of the infall of galaxies onto groups, 
we can study possible dependencies of the amplitude of
infall velocities on group properties.
It is expected that the amplitude of the infall velocities increases  
with group mass. In order to  
test this assumption we must obtain estimates of the group masses.
This can be achieved by two independent methods. The first approach is to
compute the group virial mass
which in spite of having the advantage of being a direct estimate,
is subject to significant uncertainties mainly due to small number statistics. 
A second approach is to estimate group masses by measuring the total
luminosity of group member galaxies. This method can provide a suitable 
estimate of the group mass by  assuming a constant M/L relation and is less 
unstable regarding small number statistics (Padilla et al. 2004, Eke et al. 2004).
We have considered the dependence of the 
amplitude of infall on group luminosity.
We estimate group luminosities in the $m_{Zw}$ band by adding up galaxy luminosities from
the UZC catalog that are
within $1.5$ h$^{-1}$ Mpc from the group center, with a 
relative velocity $\Delta V < 500$ km s$^{-1}$. 
This provides a rough estimate of the groups total luminosity and serves 
as an adequate procedure to deal with two subsamples of different total luminosities.
These subsamples are constructed such that they contain similar numbers of groups.

\begin{figure}
\plotone{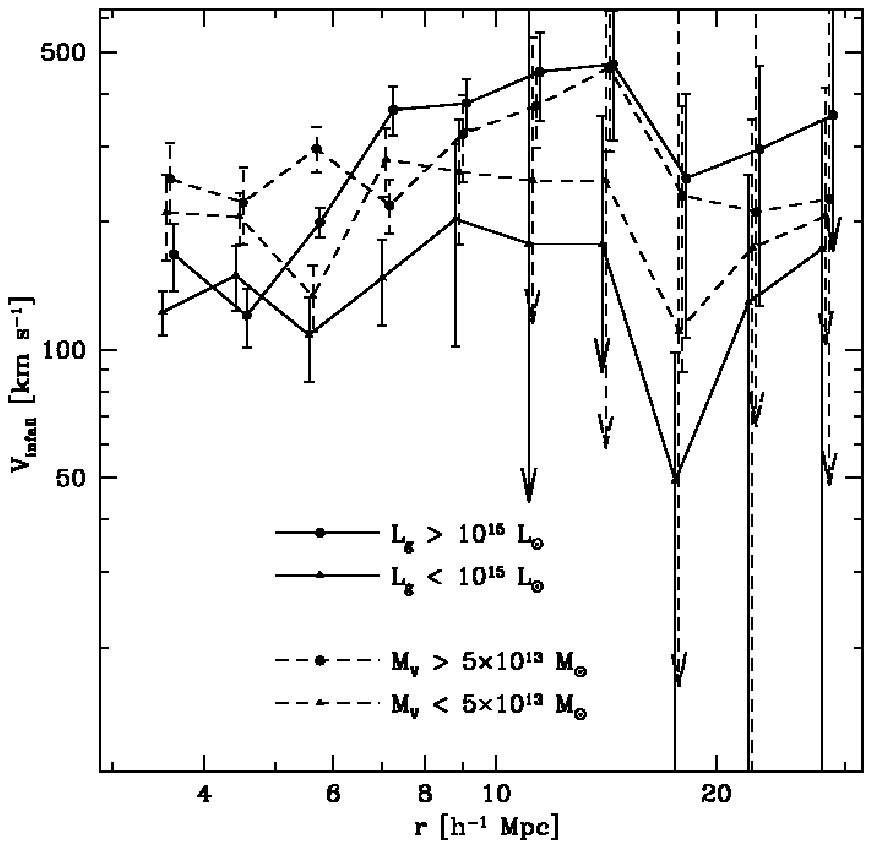}
\caption{Mean infall velocity onto samples in UZC galaxy groups. Solid lines 
indicate luminosity subsamples, and dashed lines indicate virial mass subsamples.
Errors are derived from the scatter of measurements from several mock catalogs.}
\label{fig:infmvl}
\end{figure}

In figure \ref{fig:infmvl} we show the resulting infall velocities 
derived from the two group subsamples subdivided by luminosity 
($L_{g}>10^{15}$ L$_{\odot}$ and $L_{g}<10^{15}$ L$_{\odot}$). 
As can be seen, there is a remarkable difference between the infall amplitude of both subsamples. 
We also show in figure \ref{fig:infmvl} the infall onto subsamples of
groups subdivided according to virial masses ($M_{V}>2\times10^{13}$ 
M$_{\odot}$ and $M_{V}<2\times10^{13}$ M$_{\odot}$). As in the case in which we
define our subsamples using group luminosities, 
the subdivision by mass is set so that both subsamples contain nearly equal numbers.

By inspection to figure \ref{fig:infmvl}, it can be seen that by subdividing 
the sample by luminosity, the difference in the
amplitude of infall velocities is larger than by subdividing the sample by virial mass.
This larger dependence of the amplitude of infall on group luminosity with 
respect to virial mass may reflect the fact that
the latter might be more subject to large uncertainties 
than estimates of mass derived from luminosity, in agreement with
Padilla et al. (2004) and Eke et al. (2004).

\section {Results from the Mock Catalogs}

\begin{figure}
\plotone{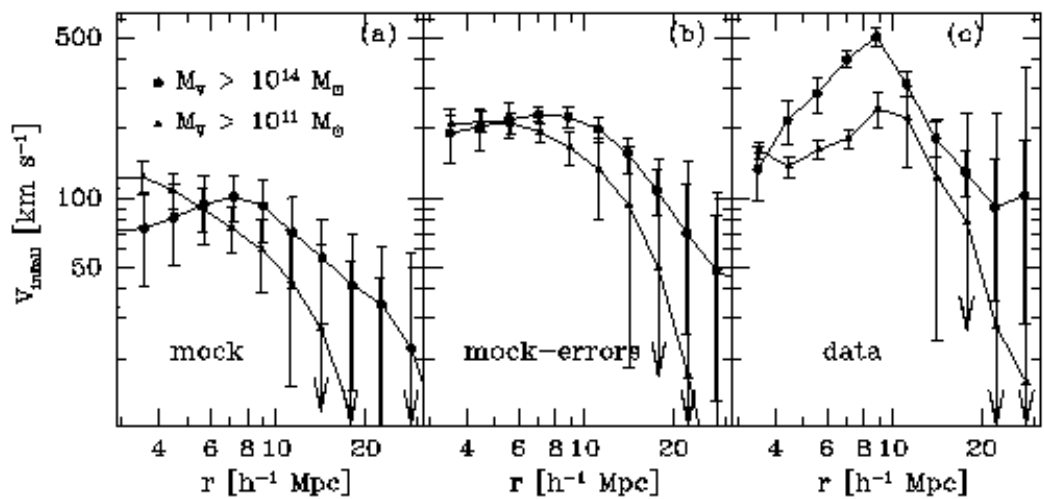}
\caption{Panel a: Mean streaming peculiar velocity toward simulated groups
for galaxies in the mock catalog. The results were
derived using the same method applied to the data. 
Triangles correspond to groups with virial mass greater than $10^{11}$ M$_{\odot}$, 
and circles correspond to groups with virial mass greater than $10^{14}$ M$_{\odot}$. 
Panel b: same as panel a, but with distances measurements errors. Panel c: Mean streaming 
peculiar velocity of CPV onto UZC groups. 
The different symbols correspond to different ranges of group virial mass,
as indicated in panel a.
Errors are derived from the scatter in several mock catalogs.}
\label{fig:vinfmock}
\end{figure}

We calculate the mean streaming peculiar velocity 
towards simulated groups for galaxies in the mock catalogs
by using the same methods that we applied to the observational data. 
In panel a) of figure \ref{fig:vinfmock} we show the mean infall motions
(averaged over seven mock catalogs) for two subsamples
of groups per mock catalog defined by two different lower limits in dynamical mass.
We find that in scales ranging from $5$ to $30$ h$^{-1}$ Mpc
the predominant motion occurs preferentially in the radial direction.  Closer
to the group centers, errors in $V_{infall}$ (calculated from the scatter
among seven mock catalog) grow larger
as virialized motions start to dominate the galaxy velocities.
At large separations, the infall signal disappears, probably
because of the effect of neighboring groups.  The results obtained
from the mock samples are at least in qualitative 
agreement with the observational data.

The resulting average infall velocities recovered from the
mock catalogs and thosse measured directly
from the simulation box are comparable.  For instance, at 
separations of $10$ h$^{-1}$Mpc, the values of infall are
$60 \pm20$km/s and $62 \pm 2$km/s for the mock and cube data,
respectively, for halos with $M>10^{11}$h$^{-1}$M$_{\odot}$.
In the case of the most massive halos with $M>10^{14}$h$^{-1}$M$_{\odot}$,
the infall velocities become $100 \pm 30$km/s and $210 \pm 3$ km/s, respectively.
As a note of caution, one should therefore bear in mind that 
the relation between infall amplitude and halo mass
seems to be slightly less step when measured from the mock catalogs,
which show consistent infall amplitudes  
for the low-mass halos and lower amplitudes
for the high-mass halos than the true underlying value.

\begin{figure}
\plotone{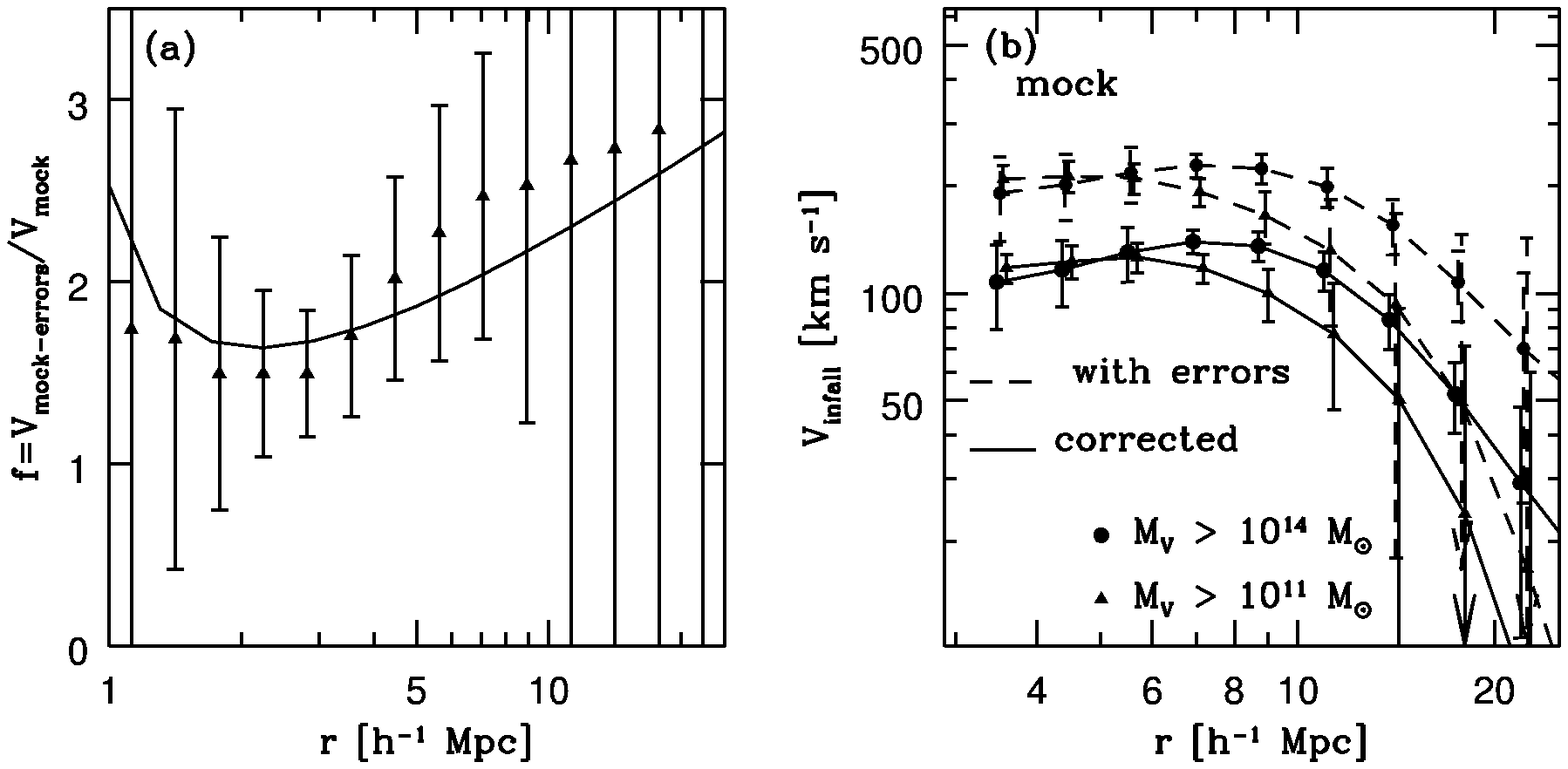}
\caption{(a) The ratio, $f$, between the infall amplitude measured from the 
mock catalogs that include distance uncertainties and the infall 
amplitude of mock catalogs without errors (triangles). 
The solid line shows a fit to the measured values of $f$ (see text). 
(b)Corrected mock infall velocities (solid lines) and infall velocities 
affected by distance uncertainties (dashed lines). Circles represent 
groups with masses greater than $10^{14}$M$_{\odot}$ and triangles  represent groups more massive 
than $10^{11}$M$_{\odot}$. 
Errors are derived from the scatter in several mock catalogs.}
\label{fig:meismi}
\end{figure}

Up to this point, peculiar velocities in the mock catalogs were not affected
by measurement errors. We apply a distance measurement error proportional to 
the distance to the galaxies similar to the one that affects the observational 
data. We consider an uncertainty of $20 \%$ in the distance measurement,
repeat the calculation of infall velocities for the mock catalogs,
and show the results in panel b of figure \ref{fig:vinfmock}.  By comparing
this panel to panel a) and panel c) which
show the infall motions for the mock catalog without errors and 
for the UZC data, respectively, it can be seen that errors in distance measurementes bring
the results from synthetic data closer to the observational data.
In order to correct for the
effect of distance uncertainties in the inferred infall velocities, 
we compare the results from the mock catalogs with and without distance errors.
We calculate the ratio $f=V_{mock-errors}/V_{mock}$, where 
$V_{mock-errors}$ is the infall amplitude of the mock catalogs 
with distance errors (see panel b in figure \ref{fig:vinfmock}) and $V_{mock}$  
is the infall amplitude from the mock catalogs without distance errors (see panel 
a in figure \ref{fig:vinfmock}). Panel a of figure \ref{fig:meismi} shows
$f$ for different scales and a fitting function of the form 
$f=a+br/(h^{-1}Mpc)+c log(r(h^{-1} Mpc))$,
which is sufficient 
to provide a good description of the ratio of observed to actual velocities. 
We find that a=2.4, b=0.125, and c=-1.94 provide a good fit to the measured values of $f$.

As it can be seen in this figure, the correction factor $f$ increases slightly 
with scale, with a mimimun value $f \simeq 1.5$ at $3 h^{-1} Mpc$ and
$f \simeq 2.8$ at $20 h^{-1} Mpc$. Thus,  in order to obtain realistic estimates, 
all the inferred infall velocities should be reduced by these values.

We note, however, that the resulting infall velocities do not depend 
significantly on the particular function adopted to fit the correction factor $f$.
Panel b in figure \ref{fig:meismi} shows the corrected mean amplitude for two 
ranges in group virial mass. We note that the lower values of the
corrected infall velocities are in suitable agreement 
with those shown in figure \ref{fig:vcube20} for mass and galaxies in the $\Lambda$ 
CDM model.

\section{Infall amplitude and mass overdensity}

In this section we assume that the linear theory prediction that relates the 
integrated mass overdensity with an infall peculiar velocity is valid
at distances greater than $5$ h$^{-1}$ Mpc. 
By doing this we can directly derive
the mean integrated mass overdensity ($\delta$) in a spherical volume centered 
on the groups.
As discussed in the previous section, the results from the mock catalogs 
indicate that the effects of distance uncertainties must be 
corrected for in order to derive the actual mean infall velocities. 
  
\begin{figure}
\plotone{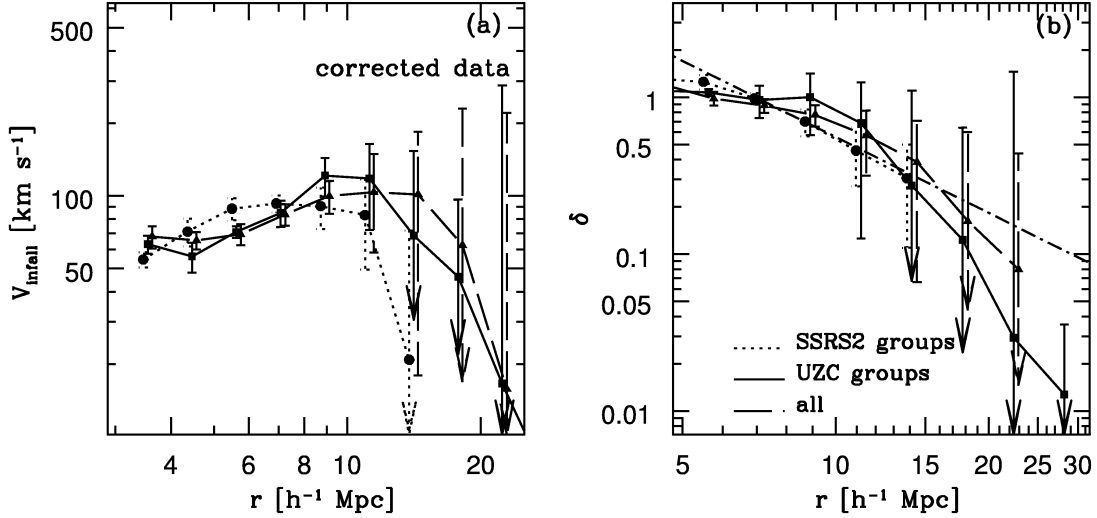}
\caption{(a)Mean infall velocity of UZC (solid line), SSRS2 (dotted line) 
and the combined set of data (dashed line), corrected for distance 
uncertainties. (b) Integrated mass 
overdensity profile derived from linear theory; in both panels the same line 
type represents the same sample. The point-dashed line in panel b indicates 
the least-squares fit to results of the combined (UZC and SSRS2) sample.
Errors are derived from the scatter in several mock catalogs.}
\label{fig:allcor}
\end{figure}

We apply the correction $f = V_{mock-errors}$/$V_{mock}$ to the infall velocities
measured from the observational data (plotted in figure  \ref{fig:vns}).
Panel (a) of figure \ref{fig:allcor} 
shows the corrected mean infall amplitudes of the samples analyzed previously.  
The mean integrated mass overdensities shown in panel (b)
 are derived from the corrected infall velocities shown in panel a calculated using
$V_{infall}(r)= -1/3 H_{0}\Omega_{0}^{0.6} r \delta (r)$, where we have assumed a 
density parameter $\Omega=0.3$ 

\begin{figure}
\plotone{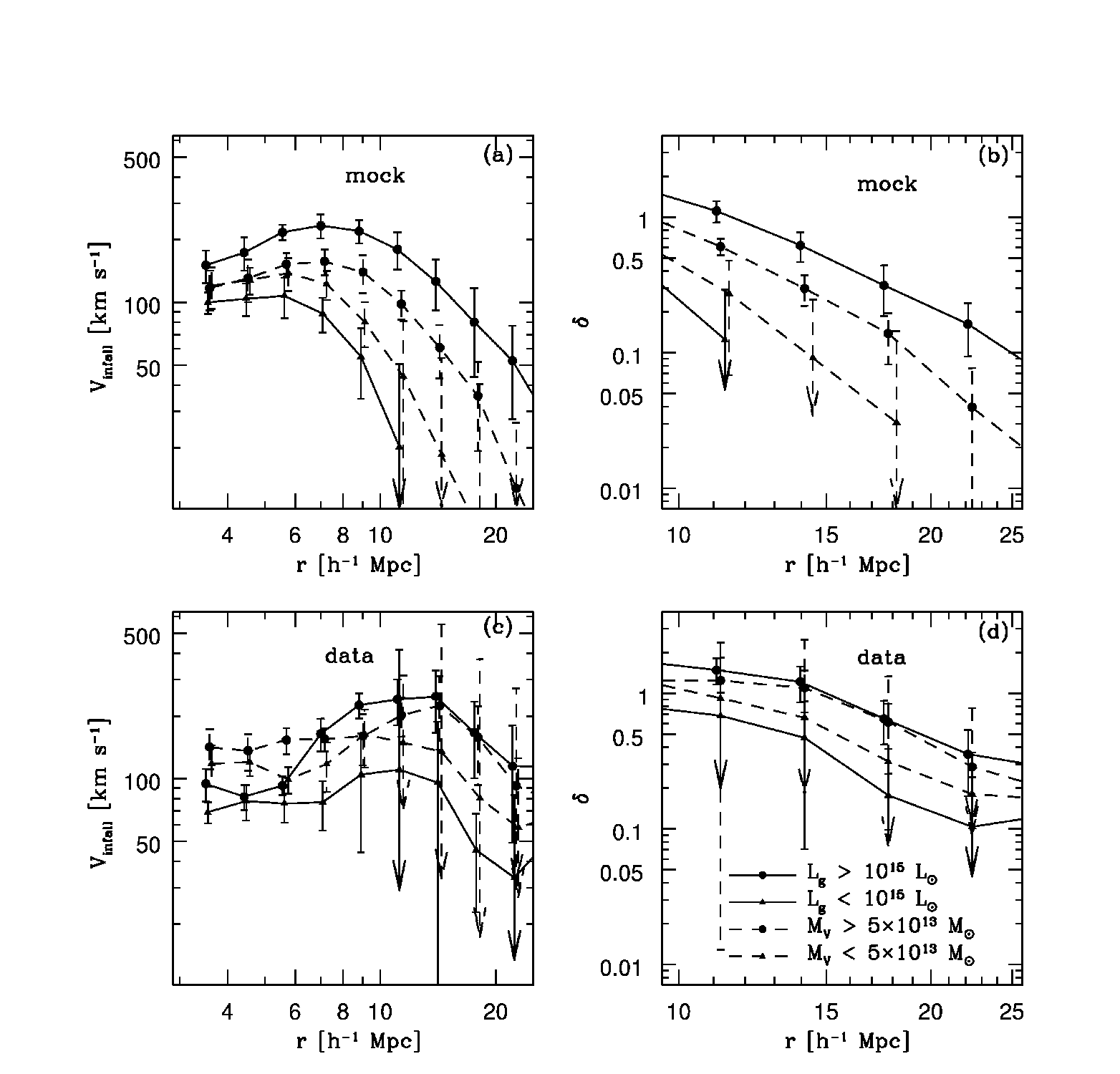}
\caption{Results from mock (a,b) and observational (c,d) data. In all four 
panels, solid lines indicate 
subsamples divided by luminosity, whereas dashed lines indicate subsamples 
divided by virial mass. 
(a) Mock infall velocity corrected by distance errors. 
(b) Integrated mass overdensity as function of radius derived from linear theory.
(c) Corrected infall velocity from observational data. (d) Integrated mass 
overdensity as function of radius derived from linear theory. 
Errors are derived from the scatter in several mock catalogs.}
\label{fig:lmock}
\end{figure}

We have also corrected for distance errors on the inferred infall velocities of galaxies 
onto groups of different luminosity and virial mass ranges.
We have accordingly applied a correction factor $f=V_{mock-errors}$/$V_{mock}$ 
to the infall velocities for which we note that the computed factors
$f$ for the different subsamples do not differ significantly.
Figure \ref{fig:lmock}  shows the corrected mean infall velocities 
and the corresponding mean integrated mass overdensities derived from linear 
theory for both the mock catalogs (panels a and b) and the observations 
(panels c and d) The mean corrected amplitude of infall velocities in the mock and
observational data in the range of scales $5$ h$^{-1}$Mpc $< r < 20$ h$^{-1}$ Mpc 
is $V\sim 100$ km s$^{-1}$  for the low-luminosity groups, 
whereas the most luminous sample reachs infall velocities as high as 
$250$ km s$^{-1}$.  In the subsamples subdivided by virial mass we have a similar 
behavior: for the least massive groups,  $V_{infall} \simeq 150$ km s$^{-1}$, whereas 
for the more massive systems, $V_{infall}\simeq 200$ km s$^{-1}$.

Table 1 lists the mean amplitude of infall velocities at $10$h $^{-1}$ Mpc, $V_{10}$,  
obtained for the UZC and the mock catalog. The integrated mass overdensities 
derived within $\sim 10$ h$^{-1}$ Mpc ($\delta_{10}$) are shown in Table 2.

\begin{deluxetable}{ccccc}
\tablecaption{Infall amplitude at $10 h^{-1} Mpc$}
\tablewidth{0pt}
\tablehead{\colhead{} &
\colhead{Observational data} & \colhead{} & \colhead{Mock data} & \colhead{}}
\startdata
\vspace{0.1cm}
& $V_{10}$[km\ s$^{-1}$]& $V_{10}$[km\ s$^{-1}$]& $V_{10}$[km\ s$^{-1}$]& $V_{10}$[km\ s$^{-1}$]\\
\hline
& crude& corrected&crude&corrected\\
\hline
$\log_{10}(M_{V}/M_{\odot})>13.7$  & $375\pm90$& $190\pm45$  & $210\pm50$& $125\pm30$ \\
$\log_{10}(M_{V}/M_{\odot})<13.7$ & $250\pm90$ & $155\pm50$ & $100\pm35$& $65\pm25$\\
$\log_{10}(L_{g}/L_{\odot})>15$  & $450 \pm 65$ & $225 \pm30$  & $350\pm50$& $210\pm30$\\
$\log_{10}(L_{g}/L_{\odot})<15$ & $180 \pm 90$& $110 \pm 60$ & $70\pm 30$ & $50\pm 20$\\
\hline
\enddata
\end{deluxetable}

\begin{deluxetable}{ccc}
\tablecaption{Integrated mass density within $10 h^{-1} Mpc$}
\tablewidth{0pt}
\tablehead{\colhead{} &
\colhead{Observational samples } & \colhead{Mock samples}}
\startdata
\vspace{0.1cm}
& $\delta_{10}$& $\delta_{10}$\\
\hline
$\log_{10}(M_{V}/M_{\odot})>13.7$  & $1.2\pm0.3$ & $0.8\pm0.2$ \\
$\log_{10}(M_{V}/M_{\odot})<13.7$ & $1.0\pm0.4$ & $0.5\pm0.2$\\
$\log_{10}(L_{g}/L_{\odot})>15$  & $1.5 \pm0.2$ & $1.4\pm0.2$\\
$\log_{10}(L_{g}/L_{\odot})<15$ & $0.8 \pm 0.4$ & $0.3\pm 0.2$\\
\hline
\enddata
\end{deluxetable}

In order to test the reliability of the estimated overdensity values 
derived from linear theory, we have compared them with the actual mass 
overdensities computed in spherical volumes of radius $10$ h $^{-1}$ Mpc  
in the simulations for different halo mass ranges. 
We obtain estimated  overdensities in the mock catalogs in
agreement with the actual values in the simulation, within the uncertainties, 
indicating the validity of the application of 
linear theory to the infall data to infer mass overdensities.

\section{Discussion}
We have analized the amplitude of the infall of galaxies onto groups and
clusters in the nearby universe.
Our study is based on the largest compilation of galaxy peculiar
velocities to date and on a well defined and controlled sample of galaxy
groups. The observational results were compared with numerical simulations,
which include galaxy formation through semianalytical models and serve
as a new test for the hierarchical models for structure formation within
the $\Lambda$CDM scenario.  The numerical simulations were also used to 
assess random and systematic errors in our measurements.

We summarize our findings as follows:
\begin{enumerate}
\item[(i)] We find a clear signature of infall of galaxies onto groups in a wide
range of scales, $5$ h$^{-1} $Mpc$ <r< 30 $ h$^{-1}$ Mpc.
The amplitude of the infall velocities is of the order of few
hundreds kilometers per second.
\item[(ii)]
There is a significant dependence of the infall amplitude on group
virial mass. For groups with $MV < 10^{13}$ M$_{\odot}$, we obtain  
$V_{infall} = 155 \pm 45$ km s$^{-1}$,
whereas for  $MV > 10^{13}$ M$_{\odot}$ a significantly larger amplitude of
infall is observed, $V_{infall }= 190 \pm 40$ km s$^{-1}$.
\item[(iii)]
In a similar fashion, the total group luminosity is a
significant parameter that influences the infall amplitude. We find that
groups with  $ L< 10^{15}$ L$_{\odot}$ have $V_{infall } 
= 110 \pm 50$ km s$^{-1}$ whereas for  $L > 10^{15}$ L$_{\odot}$
a value of $V_{infall } = 226 \pm 45$ km s$^{-1}$ is measured. We observe a larger difference 
in the amplitude of infall velocities when dividing the group sample
according to luminosity than to virial mass.
\item[(iv)]
We obtain a similar behavior for the infall of galaxies onto groups in
the numerical models and use the results from mock catalogs
to measure the random errors inherent in our measurements and to
correct the estimated infall amplitudes for systematic effects arising 
from distance measurement errors.
The results for different virial mass and luminosity thresholds are
consistent with those of the observational data, indicating that the
models are suitable to reproduce the observations.
\item[(v)]
We estimate integrated mass overdensities in spherical regions around galaxy
groups. The resulting integrated mass overdensity in spheres of radius $r$ is consistent
 with a power law  of the form $\delta \sim (r/r_1)^{-1.6}$, with
$r_1 \sim 10$ h$^{-1}$ Mpc. 
\end{enumerate}

\acknowledgments
This research was supported by grants from Agencia C\'ordoba Ciencia,
Secretar\'{\i}a de Ciencia y T\'ecnica
de la Universidad Nacional de C\'ordoba, Fundaci\'on Antorchas, and
Agencia Nacional de Promoci\'on Cientif\'{\i}ca, Argentina.
NP was supported by a FONDECYT grant no. 3040038, Chile.
This work was partly supported by the ESO grant at PUC, Chile,
and NSF grant AST-0307396 to RG.

\end{document}